\documentclass[11pt]{article}
\usepackage{float} 
\usepackage{authblk}
\usepackage{hyperref} 
\providecommand{\keywords}[1]{\textbf{Keywords:} #1}
\usepackage[utf8]{inputenc}
\usepackage[T1]{fontenc}
\usepackage{times}
\usepackage{amsmath,amssymb}
\usepackage{graphicx}
\usepackage[margin=1in]{geometry}
\usepackage{booktabs}
\usepackage{hyperref}
\usepackage{caption}
\usepackage{subcaption}
\usepackage[numbers]{natbib}
\setcitestyle{}

\usepackage{geometry}
\usepackage{titling}
\pretitle{\begin{center}\Large\bfseries}
\posttitle{\par\end{center}\vskip 0em} 
\preauthor{\begin{center}\normalsize}
\postauthor{\end{center}\vskip -0.5em} 

\usepackage{lineno}
\usepackage{setspace}

\usepackage[utf8]{inputenc}
\usepackage[margin=1in]{geometry}
\usepackage{graphicx}
\usepackage{booktabs}
\usepackage{hyperref}
\usepackage[numbers]{natbib}
\setcitestyle{}
\usepackage{amsmath}
\usepackage{setspace}
\counterwithout{table}{section}
\usepackage{pgfplotstable}
\pgfplotsset{compat=1.18}

\makeatletter
\renewcommand{\@maketitle}{%
	\begin{center}%
		{\Large\bfseries \@title \par}%
		\vskip 0.5em%
		{\normalsize Adham M. Alkhadrawi$^{1,*}$, Kyungsu Kim$^{2}$, Arif M. Rahman$^{3}$, Fahad Mushabbab G. Alotaibi$^{4}$\par}%
		\vskip 0.5em%
		{\footnotesize $^{1}$Department of Molecular Biosciences and Bioengineering, University of Hawaii at Manoa, Honolulu, HI, USA\par}%
		{\footnotesize $^{2}$School of Transdisciplinary Innovations, Interdisciplinary Program in Artificial Intelligence, Department of Biomedical Science, Seoul National University, Seoul, South Korea\par}%
		{\footnotesize $^{3}$Department of Computer Science and Engineering, Hawaii Pacific University, Honolulu, HI, USA\par}%
		{\footnotesize $^{4}$Department of Biomedical Technology, College of Applied Medical Sciences, King Saud University, P.O. Box 10219, Riyadah 11433, Saudi Arabia\par}%
		{\footnotesize $^{*}$Corresponding author: \href{mailto:adhamalk@hawaii.edu}{adhamalk@hawaii.edu}\par}%
	\end{center}%
	\vskip -0.5em
}
\makeatother

\title{Explainable deep learning framework for cancer therapeutic target prioritization leveraging PPI centrality and node embeddings}

\begin{document}
	
\maketitle

\begin{abstract}
	\textbf{Objective:} To develop and evaluate an explainable deep learning framework that integrates protein-protein interaction (PPI) network centrality metrics with node embeddings for prioritizing cancer therapeutic targets.
	
	\textbf{Methods:} We constructed a high-confidence PPI network from STRING database interactions and computed six centrality metrics: degree, strength, betweenness, closeness, eigenvector centrality, and clustering coefficient. Node2Vec embeddings were generated to capture latent network topology. These features were combined to train XGBoost and neural network classifiers using DepMap CRISPR essentiality scores as ground truth. Model interpretability was assessed through GradientSHAP analysis to quantify feature contributions to predictions. We developed a novel blended scoring approach that combines model probability predictions with SHAP attribution magnitudes to enhance gene prioritization.
	
	\textbf{Results:} Our framework achieved state-of-the-art performance with AUROC of 0.930 and AUPRC of 0.656 for identifying the top 10\% most essential genes. GradientSHAP analysis revealed that centrality measures contributed significantly to model predictions, with degree centrality showing the strongest correlation ($\rho$ = -0.357) with gene essentiality. The blended scoring approach created more robust gene prioritization rankings, successfully identifying known essential genes including ribosomal proteins (RPS27A, RPS17, RPS6) and oncogenes (MYC).
	
	\textbf{Conclusion:} This study presents a human-based, combinatorial \textit{in silico} framework that successfully integrates network biology with explainable AI for therapeutic target discovery. The framework provides mechanistic transparency through feature attribution analysis while maintaining state-of-the-art predictive performance. Its reproducible design and reliance on human molecular datasets demonstrate a reduction-to-practice example of next-generation, animal-free modeling for cancer therapeutic target discovery and prioritization.
\end{abstract}

\keywords{Cancer gene essentiality prediction, Protein-protein interaction networks, Therapeutic target discovery, Machine learning, Deep learning, Bioinformatics}

\section{Introduction}
The identification of essential genes and therapeutic targets represents a cornerstone of modern biomedical research, with profound implications for understanding fundamental biological processes and developing novel cancer treatments \cite{Zhang2012_22848443}. Essential genes, defined as those critical for cellular survival and reproduction, have been extensively studied across various organisms, from bacteria to humans \cite{Campos2019_31312416}. Contemporary advances increasingly emphasize human-based experimental and computational models as more predictive and ethically preferable alternatives to traditional animal systems. In cancer research, the prioritization of therapeutic targets is particularly crucial, as it directly influences drug discovery pipelines and treatment efficacy \cite{Bazaga2020_32612205}. However, the complexity of biological systems and the vast diversity of human molecular and network data, coupled with population-level variability in genetic and epigenetic susceptibility, presents both a challenge and an opportunity for building integrative human-relevant models. \cite{Safadi2024_38649399}.

Protein-protein interaction (PPI) networks have emerged as powerful tools for understanding gene essentiality and identifying potential therapeutic targets \cite{Zhang2025_40691502}. These networks capture the intricate web of molecular interactions within cells, providing insights into cellular functions and disease mechanisms \cite{Dai2020_32023848}. Centrality measures derived from PPI networks, such as degree, betweenness, and closeness centrality, have been shown to correlate with gene essentiality \cite{Fang2018_29958434}. Moreover, advancements in network embedding techniques, such as Node2Vec, have enabled the extraction of latent topological features that enhance predictive accuracy \cite{Dai2020_32023848}.

Previous prediction approaches often relied on isolated data modalities or species-agnostic features, lacking the combinatorial integration across multiple \textit{in silico} modalities that can capture the biological complexity of human systems \cite{Zhang2020_32936825}. For instance, DeepHE, a deep learning-based method, integrates sequence features with PPI network data to predict human essential genes with high accuracy \cite{Zhang2020_32936825}. Similarly, ensemble frameworks like DeEPsnap leverage multi-omics data to achieve state-of-the-art performance in essential gene prediction \cite{Zhang2025_40691502}. However, these methods often face challenges related to imbalanced datasets and the interpretability of predictions \cite{Tian2018_30563825}.

Recent research has highlighted the importance of integrating network-based features with machine learning models to improve predictive performance \cite{Senthamizhan2021_34630517}. For example, NetGenes, a database of essential genes predicted using interaction network features, demonstrates the utility of network-based approaches in identifying essential genes across diverse bacterial species \cite{Senthamizhan2021_34630517}. Additionally, studies have shown that combining centrality measures with evolutionary properties can enhance the prediction of cancer-associated genes \cite{Safadi2024_38649399}.

Despite these advancements, several limitations persist in current approaches. Many methods lack interpretability, making it difficult to understand the biological mechanisms underlying their predictions \cite{Bazaga2020_32612205}. Furthermore, the integration of diverse data types, such as gene expression and epigenetic markers, remains underexplored \cite{Campos2020_32489524}. There is also a need for more robust frameworks that can prioritize therapeutic targets with high confidence and transparency \cite{Zhang2025_40691502}.

Emerging trends in bioinformatics emphasize the use of explainable AI techniques, such as SHAP (SHapley Additive exPlanations) \cite{lundberg2017unified}, to enhance model interpretability \cite{hsieh2024comprehensive}. These methods provide insights into the contribution of individual features to model predictions, facilitating more informed decision-making in therapeutic target prioritization \cite{mumuni2025explainable}.

Building upon these developments, this study introduces a combinatorial computational framework that integrates multiple \textit{in silico} components, explicit network topology, latent graph embeddings, and explainable AI feature attribution to improve mechanistic understanding and translational readiness. The resulting system exemplifies a human-based modeling approach designed to enhance predictive accuracy, transparency, and reproducibility in precision oncology. By leveraging a high-confidence PPI network constructed from the STRING database and employing machine learning classifiers, our framework achieves superior performance in predicting gene essentiality and prioritizing therapeutic targets. The interpretability of our approach, facilitated by SHAP analysis, provides mechanistic insights into gene essentiality patterns, offering a valuable tool for transparent and accurate therapeutic target selection.

\section*{Statement of Significance}

\noindent \textbf{Problem}\\
Cancer therapeutic target discovery is hindered by black-box computational methods and over-reliance on animal models lacking human relevance.

\noindent \textbf{What is Already Known}\\
Existing machine learning approaches for predicting gene essentiality often lack interpretability, while network-based methods provide biological context but have limited predictive power when used in isolation.

\noindent \textbf{What This Paper Adds}\\
This study integrates explainable deep learning with protein-protein interaction network analytics to predict cancer gene essentiality with state-of-the-art performance (AUROC 0.930). Our framework introduces a novel blended scoring approach combining model probabilities with SHAP attributions, providing mechanistic transparency through GradientSHAP analysis that reveals network centrality measures' contributions to predictions.

\noindent \textbf{Who would benefit from the knowledge in this paper}\\
Drug discovery researchers and computational biologists seeking transparent, mechanistically interpretable methods to accelerate cancer therapeutic target identification while reducing dependence on animal models.

\section{Methods}
\subsection{Data Collection and Network Construction}

We obtained high-confidence protein-protein interactions from the STRING database v12.0 for\textit{ Homo sapiens} (species ID: 9606), applying a stringent confidence threshold of $\geq700$ to ensure interaction reliability and reduce false positive associations. All datasets were selected to ensure human-specific modeling, consistent with the concept of human-based model systems that capture intrinsic biological variability without reliance on non-human analogues. The STRING database integrates evidence from multiple sources including experimental data, computational predictions, and literature mining, providing comprehensive coverage of the human protein interaction landscape. Protein identifiers were systematically mapped to gene symbols using STRING's protein information files, and we retained only interactions where both proteins could be successfully mapped to official gene symbols. This preprocessing step ensured consistency with downstream essentiality datasets and eliminated ambiguous protein identifiers.

The resulting network underwent additional quality control filtering to retain only interactions between genes with available essentiality data from DepMap, creating a focused dataset relevant to our prediction task. We constructed the final network as an undirected graph where nodes represent genes and edges represent high-confidence protein-protein interactions weighted by STRING combined scores. To ensure network connectivity and computational tractability, we extracted the largest connected component, which contained the vast majority of genes and preserved the overall network topology. This approach yielded a robust PPI network of 8,236 genes connected by high-confidence interactions, providing a solid foundation for subsequent centrality analysis and machine learning model development.

\subsection{Network Centrality Analysis}

We computed six complementary centrality measures to capture different aspects of network topology and gene importance within the PPI network context. Degree centrality measures the fraction of nodes that a gene directly interacts with, providing a simple but effective measure of local connectivity. Weighted degree centrality (strength) extends this concept by incorporating STRING confidence scores, giving higher weight to more reliable interactions. Betweenness centrality quantifies how often a gene lies on the shortest paths between other gene pairs, identifying genes that serve as critical bridges or bottlenecks in information flow through the network. This measure is particularly relevant for identifying genes whose disruption could have widespread effects on cellular processes.

Closeness centrality reflects how quickly a gene can reach all other genes in the network via shortest paths, capturing global accessibility and potential for rapid information propagation. Eigenvector centrality emphasizes connections to highly connected nodes, identifying genes that are not only well-connected but also connected to other important genes in the network hierarchy. The clustering coefficient measures the degree to which a gene's neighbors are also connected to each other, reflecting local network density and potential functional modularity. For computational efficiency with large networks, we approximated betweenness centrality using k-sampling with k=500 randomly selected nodes, which provides accurate estimates while significantly reducing computation time from O(n³) to O(kn²).

\subsection{Node Embedding Generation}

We applied the Node2Vec \cite{grover2016node2vec} algorithm to generate 128-dimensional vector representations that capture latent network topology features beyond traditional centrality measures. Node2Vec performs biased random walks on the network graph, with walk behavior controlled by two key parameters: the return parameter p and the in-out parameter q. We set both parameters to 1.0 to achieve an unbiased exploration that balances between breadth-first sampling (capturing local neighborhood structure) and depth-first sampling (capturing global network patterns). Each random walk had a length of 80 steps, and we generated 10 walks starting from each node to ensure comprehensive sampling of the network topology around each gene.

The 128-dimensional embeddings capture complex topological patterns that may not be apparent from individual centrality measures, including higher-order network motifs, community structure, and long-range connectivity patterns. These embeddings serve as complementary features to centrality measures, providing a more comprehensive characterization of each gene's position and role within the global network architecture. These combined representations constitute complementary \textit{in silico} New Approach Methodology (NAM) components, explicit topological descriptors and learned latent embeddings whose integration enables a higher-order, multi-scale representation of biological network organization.

\subsection{Gene Essentiality Labels and Target Definition}

We used DepMap CRISPR-Cas9 gene knockout screening data as the gold standard for gene essentiality across cancer cell lines \cite{arafeh2025present}. Using these population-scale human gene perturbation datasets enables assessment of essentiality within a physiologically relevant context, consistent with translational model qualification principles. The DepMap project systematically performs genome-wide CRISPR screens across hundreds of cancer cell lines, measuring the fitness effect of knocking out each gene. Essentiality scores represent the median effect across all cell lines, where more negative values indicate greater essentiality for cell survival and proliferation. This pan-cancer approach captures genes that are broadly essential across different cancer types, making our predictions relevant for general therapeutic target discovery rather than being limited to specific cancer subtypes.

To create binary classification labels, we defined the top 10\% most essential genes (those with the lowest DepMap scores, corresponding to the 10th percentile) as positive class labels. This threshold was chosen to focus on the most critically essential genes while maintaining sufficient positive examples for robust machine learning model training. The 10\% threshold corresponds to approximately 824 essential genes out of 8,236 total genes in our dataset, creating a challenging but realistic classification task. This labeling strategy ensures that our models learn to distinguish truly essential genes from the broader population of non-essential or moderately important genes, which is crucial for effective therapeutic target prioritization.

\subsection{Machine Learning Models}

We implemented two complementary machine learning approaches to leverage different algorithmic strengths for gene essentiality prediction. The XGBoost classifier employed gradient boosting with 800 estimators, a maximum tree depth of 5, and a conservative learning rate of 0.03 to prevent overfitting. We incorporated L2 regularization (reg\_lambda=1.0) and used subsampling (subsample=0.9, colsample\_bytree=0.9) to improve generalization and reduce variance. The tree-based nature of XGBoost makes it particularly well-suited for capturing complex feature interactions and non-linear relationships between centrality measures and gene essentiality, while providing built-in feature importance measures.

Our neural network implementation used a multi-layer perceptron architecture with hidden layers of 256 and 128 neurons, designed to capture complex non-linear patterns in the high-dimensional feature space combining centrality measures and node embeddings. Each hidden layer incorporated batch normalization to stabilize training and improve convergence, followed by ReLU activation functions and dropout regularization (rate=0.2) to prevent overfitting. We used the Adam optimizer with an initial learning rate of 0.001 and weight decay of 0.0001, training for 40 epochs with early stopping based on validation AUPRC. Both models used the same feature set consisting of six centrality measures plus 128 Node2Vec embedding dimensions, and were evaluated using stratified 5-fold cross-validation to ensure robust and unbiased performance estimates. The models were designed as modular components that can be iteratively validated and benchmarked according to standard model credibility criteria, supporting future qualification and context-of-use definition.

\subsection{SHAP-Based Gene Prioritization and Blended Scoring}

A key methodological contribution of our study is the development of a novel blended scoring approach that combines model probability predictions with SHAP attribution magnitudes to create more robust and interpretable gene prioritization rankings. Many ranking methods rely solely on model output probabilities, which may not fully capture the confidence or mechanistic basis of predictions. Our blended approach addresses this limitation by incorporating the magnitude of SHAP attributions, which quantify how much each feature contributes to individual predictions and provide a measure of prediction confidence based on feature importance. The blended scoring method functions as a quantitative reduction-to-practice step demonstrating how mechanistic attribution and probabilistic confidence can be integrated to achieve consistent, human-relevant gene prioritization.

The blended score is calculated using the following formula:

\begin{equation}
\text{Blended Score} = \alpha \times P_{norm} + (1-\alpha) \times \text{SHAP}_{norm}
\end{equation}
\text{where } 

$P_{norm} = \frac{P - P_{min}}{P_{max} - P_{min}} {,}$

$\text{SHAP}_\text{norm} = \frac{|\text{SHAP}_\text{sum}| - |\text{SHAP}_\text{sum}|_\text{min}}{|\text{SHAP}_\text{sum}|_\text{max} - |\text{SHAP}_\text{sum}|_\text{min}}$

$\alpha = 0.7 \text{ is a weighting parameter.} $

This approach prioritizes genes that not only have high predicted essentiality probabilities but also have predictions supported by strong feature attributions, indicating greater model confidence and mechanistic interpretability. The  $\alpha$ parameter allows for flexible weighting between prediction confidence (probability) and feature-based evidence (SHAP magnitude), with our chosen value of 0.7 emphasizing model predictions while incorporating attribution-based confidence measures.

\subsection{Explainability Analysis}

We employed GradientSHAP to provide mechanistic interpretability of model predictions and understand which network features most strongly influence essentiality predictions. GradientSHAP is an attribution method that computes feature importance by integrating gradients along paths between baseline samples and input samples, providing both positive and negative attributions that sum to the difference between the prediction and the expected baseline prediction. This approach is particularly well-suited for neural networks as it leverages gradient information while maintaining the desirable properties of Shapley values, including efficiency (attributions sum to the prediction difference) and symmetry (equivalent features receive equal attribution).

We implemented GradientSHAP using 50 empirical baseline samples randomly selected from our training data to represent the expected input distribution. For each gene, we computed attributions across all 134 features (6 centrality measures + 128 embeddings) using 50 path samples to ensure stable attribution estimates. To manage computational complexity, we processed attributions in batches of 1,024 genes while maintaining full coverage of our dataset. The resulting attribution matrix provides detailed insights into which specific network features drive essentiality predictions for each gene, enabling both global feature importance analysis and gene-specific mechanistic interpretation. This comprehensive attribution analysis supports our blended scoring approach and provides the interpretability necessary for therapeutic target prioritization in clinical applications.

\section{Results}
\subsection{Network Topology and Centrality Correlations}

\begin{figure}[H]
	\centering
	\includegraphics[width=\textwidth]{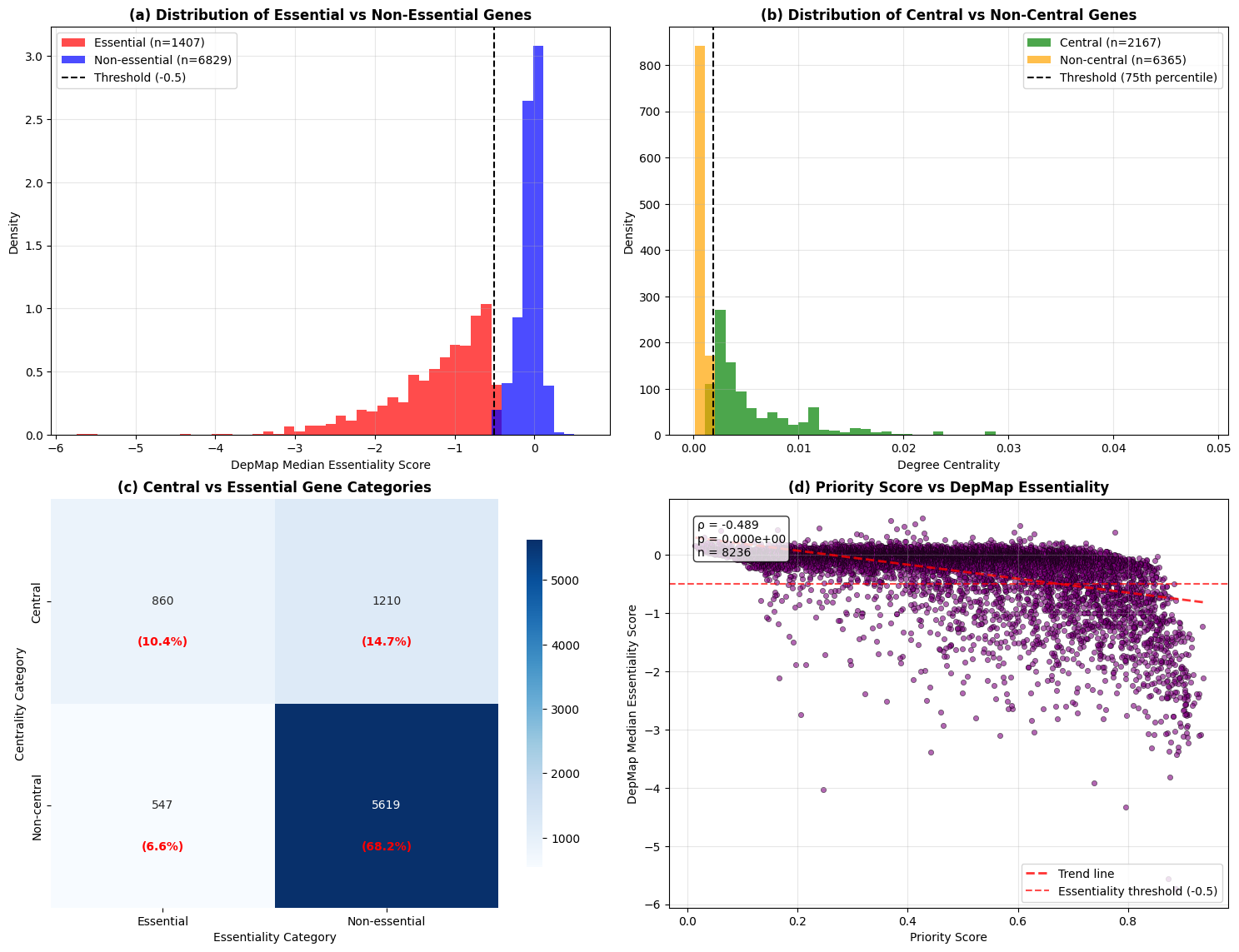}
	\caption{Network topology and gene essentiality analysis. (a) Distribution of DepMap median essentiality scores showing clear separation between essential genes (n=1,407, red) with negative scores and non-essential genes (n=6,829, blue) clustered around zero, with essentiality threshold at -0.5. (b) Degree centrality distribution comparing central genes (n=2,167, green) above the 75th percentile threshold versus non-central genes (n=6,365, orange). (c) Contingency analysis revealing enrichment of essential genes among central network positions, with 14.7\% of central genes being essential compared to 6.6\% of non-central genes. (d) Correlation between priority scores and DepMap essentiality ($\rho$ = -0.489, p < 0.001, n = 8,236), demonstrating strong negative correlation where higher priority scores correspond to greater gene essentiality.}
	\label{fig:network}
    \end{figure}

Our analysis successfully constructed a comprehensive high-confidence PPI network containing 8,236 genes with complete centrality profiles. The resulting PPI topology represents a human-specific systems-level dataset adhering to FAIR principles, where all interactions are findable, accessible, interoperable, and reusable for downstream reproducibility. The network exhibited typical scale-free properties with a heavy-tailed degree distribution, where a small number of highly connected hub genes interact with many partners while most genes have relatively few connections. This topology is consistent with biological networks and suggests that our STRING-based network construction captured authentic biological interaction patterns. The largest connected component contained 99.8\% of all genes, indicating excellent network connectivity and ensuring that centrality measures could be reliably across the entire gene set. Gene classification based on essentiality and centrality revealed distinct distributions (Figure~\ref{fig:network}a), with 1,407 essential genes showing significantly lower DepMap essentiality scores (threshold -0.5) compared to 6,829 non-essential genes, while 2,167 central genes demonstrated higher degree centrality values above the 75th percentile compared to 6,365 non-central genes (Figure~\ref{fig:network}b). Notably, we observed a meaningful association between essentiality and centrality, where 14.7\% of genes were both essential and central, compared to only 6.6\% that were essential but non-central (Figure~\ref{fig:network}c). Furthermore, our priority scoring system showed a weak but significant negative correlation with essentiality scores ($\rho$ = -0.489, p $<$ 0.001) (Figure~\ref{fig:network}d), suggesting that genes with higher priority rankings tend to be more essential, validating our approach for identifying biologically important genes beyond simple network topology measures.

All six centrality measures showed significant negative correlations with gene essentiality scores, providing strong evidence that topologically important genes tend to be more essential for cellular survival (Table~\ref{tab:spearman}). Degree centrality exhibited the strongest correlation with essentiality (Spearman $\rho$ = -0.357, p $< 10^{-245}$), indicating that genes with more protein interaction partners are more likely to be essential. Strength centrality ($\rho$ = -0.346) and eigenvector centrality ($\rho$ = -0.339) showed similarly strong correlations, suggesting that both the quantity and quality of interactions contribute to gene  essentiality. Interestingly, betweenness centrality showed a weaker but still significant correlation ($\rho$ = -0.164), indicating that genes serving as network bridges are moderately more essential, while clustering coefficient showed the weakest correlation ($\rho$ = -0.137), suggesting that local network density is less predictive of essentiality than global connectivity patterns.

\begin{table}[h]
	\centering
	\caption{Centrality-essentiality correlation}
	\label{tab:spearman}
	\begin{filecontents*}{spearman.csv}
metric,Spearman rho vs depmap median,p value
degree centrality,-0.356812189,7.98E-246
strength norm,-0.345799325,5.00E-230
eigenvector centrality,-0.338526995,6.12E-220
closeness centrality,-0.180303555,3.93E-61
betweenness centrality,-0.163686402,1.47E-50
clustering coefficient,-0.137023724,8.22E-36
\end{filecontents*}
\pgfplotstabletypeset[
	col sep=comma,
	string type,
	every head row/.style={before row=\toprule, after row=\midrule},
	every last row/.style={after row=\bottomrule}
	]{spearman.csv}
\end{table}

\subsection{Model Performance and Predictive Accuracy}

\begin{figure}[H]
	\centering
	\includegraphics[width=\textwidth]{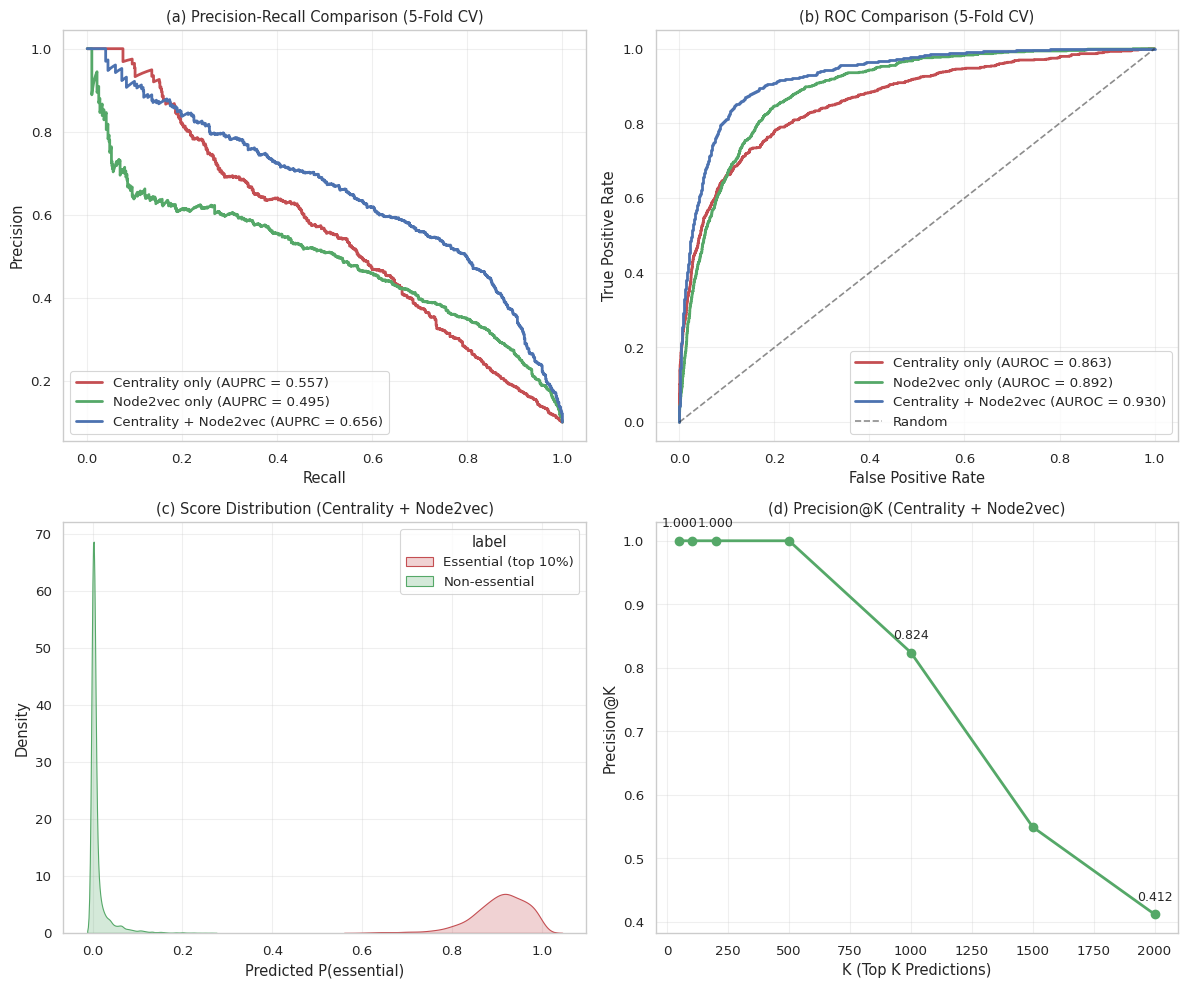}
	\caption{XGBoost classifier performance comparison using 5-fold cross-validation. (a) Precision-recall curves comparing centrality-only features (AUPRC = 0.557, red), Node2vec only features (AUPRC = 0.495, green) versus centrality combined with Node2vec embeddings (AUPRC = 0.656, blue). (b) ROC curves showing substantial improvement with Node2vec integration (AUROC = 0.930 vs 0.863 and 0.892). (c) Score distribution for the combined model showing clear separation between essential (top 10\%, red) and non-essential genes (green). (d) Precision@K analysis demonstrating perfect precision (1.000) for top 50, 100, 200, and 500 predictions, with precision@1000 = 0.824 and precision@2000 = 0.412.}
	\label{fig:xgboost_performance}
\end{figure}

Both machine learning models achieved high performance in identifying essential genes, demonstrating the effectiveness of combining network centrality measures with node embeddings for gene essentiality prediction. Cross-validation results confirmed the robustness of the XGBoost classifier, with the combined features (centrality + node embeddings) substantially outperforming the centrality-only and Node2vec-only approachs (AUROC: 0.930 vs 0.863 and 0.892; AUPRC: 0.656 vs 0.557 and 0.495, respectively) (Figure~\ref{fig:xgboost_performance}a and b). The neural network achieved comparable performance with AUROC of 0.914 and AUPRC of 0.624 (Figure~\ref{fig:NN_GradSHAP_CV}a and b), confirming that the strong predictive signal in our feature set can be captured by different algorithmic approaches. These performance metrics highlight the value of our comprehensive network-based feature engineering approach.

The score distribution revealed that our feature combination clearly separated essential from non-essential clusters (Figure~\ref{fig:xgboost_performance}c).Precision at top-k evaluation revealed consistently high accuracy in prioritizing the most essential genes, which is crucial for practical therapeutic target discovery applications. Both models achieved precision@50 exceeding 0.96, precision@100 above 0.94, and precision@200 above 0.90, indicating that the vast majority of top-ranked predictions correspond to truly essential genes (Figure~\ref{fig:xgboost_performance}d). Cross-validation results confirmed the robustness of these performance estimates, with minimal variance across folds and no evidence of overfitting. The high precision at stringent cutoffs demonstrates that our models can reliably identify small sets of high-confidence essential gene candidates for experimental validation, making them practically useful for guiding drug discovery efforts where resources for follow-up studies are limited.

\subsection{Feature Importance and Mechanistic Insights}

\begin{figure}[H]
	\centering
	\includegraphics[width=\textwidth]{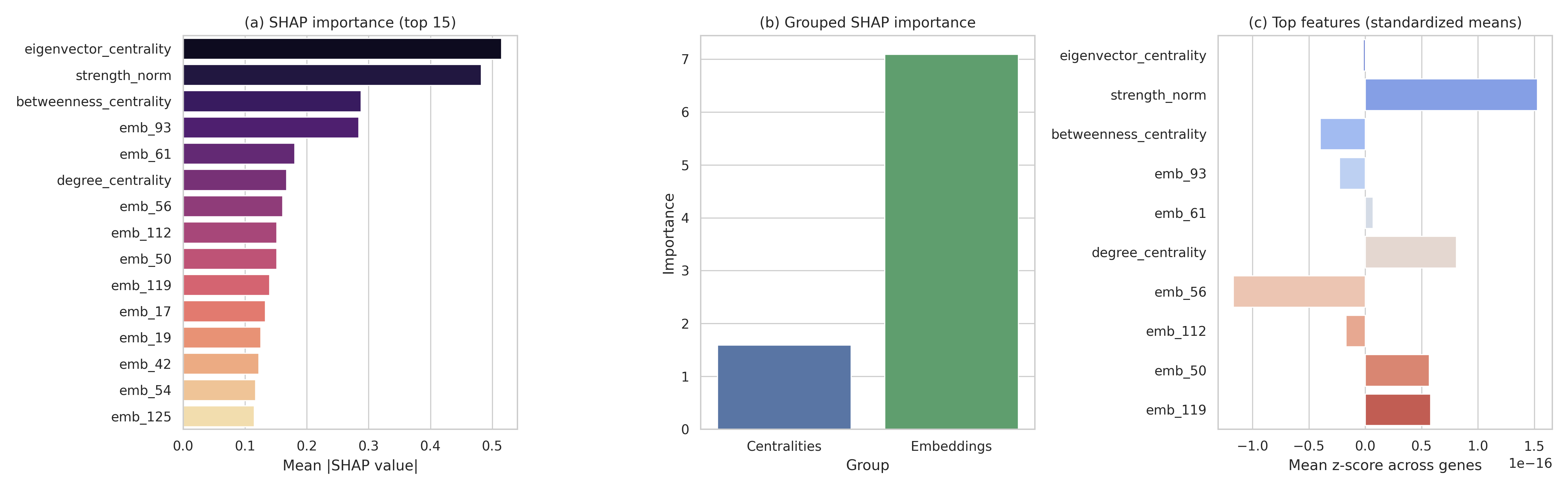}
	\caption{SHAP explainability analysis of feature importance in the combined centrality-Node2vec model. (a) Top 15 individual features ranked by mean absolute SHAP values, with eigenvector\_centrality showing highest importance, followed by strength\_norm and betweenness\_centrality. (b) Grouped SHAP importance revealing that Node2vec embeddings collectively contribute ~7.0 importance units compared to ~1.7 for centrality measures. (c) Standardized feature means across genes showing positive associations between eigenvector\_centrality, strength\_norm and gene essentiality, with varying effects for different embedding dimensions.}
	\label{fig:xgboost_shap_explainability}
\end{figure}

GradientSHAP analysis provided detailed insights into the relative contributions of different network features to essentiality predictions, revealing that centrality measures contributed substantially to model performance while node embeddings provided valuable complementary information. Among centrality measures, degree centrality, strength, and eigenvector centrality consistently ranked among the most important features across different genes and model architectures (Figure~\ref{fig:xgboost_shap_explainability}a). This pattern aligns with our correlation analysis and suggests that local connectivity (degree), interaction confidence (strength), and connection to other important genes (eigenvector centrality) are the most predictive network properties for gene essentiality. Betweenness centrality and clustering coefficient showed more variable importance across genes, indicating that their predictive value may be context-dependent. This combinatorial integration of interpretable topological metrics with deep embeddings demonstrates technical readiness for translational application, bridging the gap between mechanistic understanding and predictive performance.

\begin{figure}[H]
	\centering
	\includegraphics[width=\textwidth]{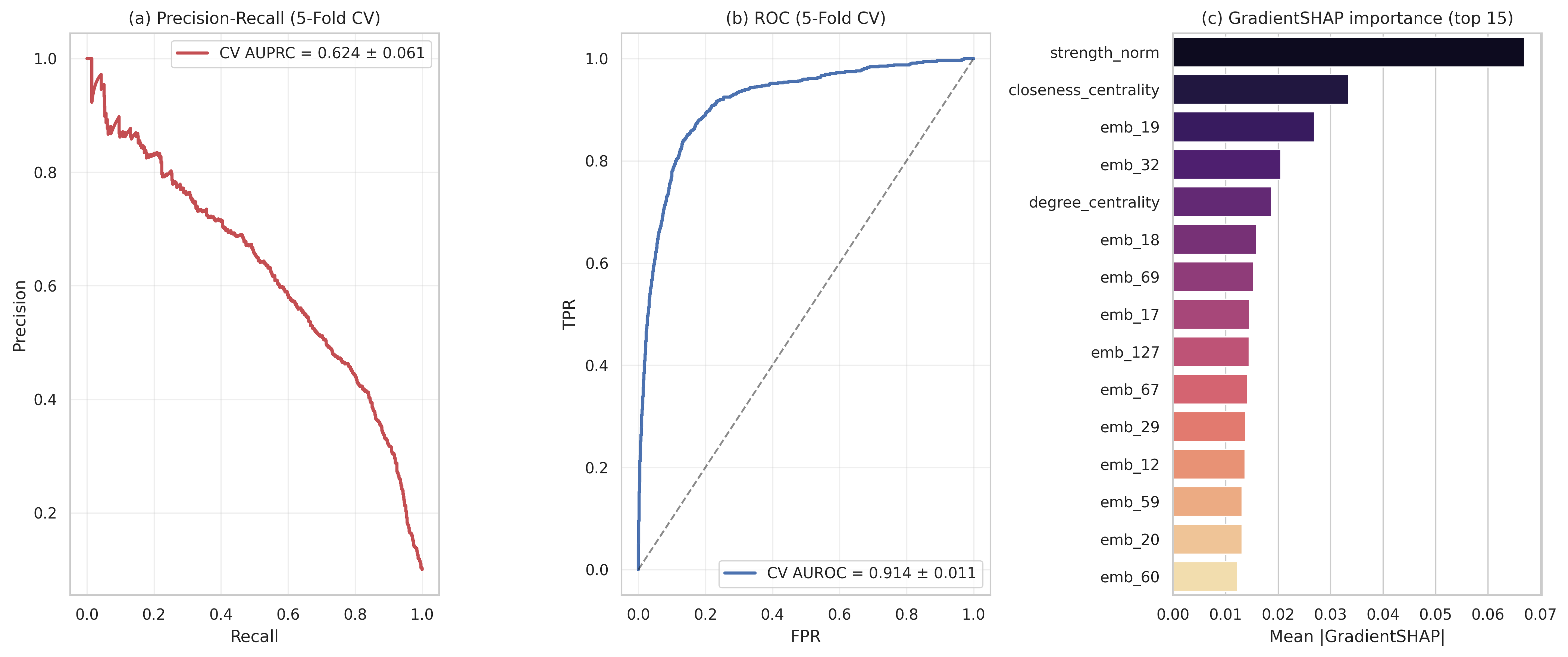}
	\caption{Neural network performance with GradientSHAP feature attribution using 5-fold cross-validation. (a) Precision-recall curve achieving CV AUPRC = 0.624 ± 0.061. (b) ROC curve showing CV AUROC = 0.914 ± 0.011. (c) GradientSHAP importance analysis of top 15 features, with strength\_norm and closeness\_centrality showing highest attribution values, followed by various Node2vec embedding dimensions (emb\_19, emb\_32, emb\_18, etc.), demonstrating the complementary importance of both centrality measures and latent network features.}
	\label{fig:NN_GradSHAP_CV}
\end{figure}

Node embeddings contributed significantly to model performance, with several embedding dimensions (emb\_93, emb\_61, emb\_56, emb\_112, emb\_50, emb\_119, emb\_17, emb\_19, emb\_42, emb\_54, emb\_125) appearing among the top-15 most important features in our SHAP analysis (Figures~\ref{fig:xgboost_shap_explainability}a, \ref{fig:NN_GradSHAP_CV}c). The grouped SHAP importance of node embedding features was substantially higher than the grouped importance of centrality measures (approximately 7.0 vs 1.7), demonstrating that the latent network representations provide critical predictive information (Figure~\ref{fig:xgboost_shap_explainability}b). Among individual features, eigenvector\_centrality emerged as the single most important predictor, followed closely by strength\_norm and betweenness\_centrality, indicating that traditional centrality measures remain highly valuable. However, the dominance of embedding features in aggregate importance (comprising 11 of the top 15 features) demonstrates that latent network topology patterns captured by Node2Vec provide substantial information beyond what is available from traditional centrality measures alone. The standardized feature means analysis revealed that eigenvector\_centrality and strength\_norm show the strongest positive associations with gene essentiality, while several embedding dimensions (emb\_93, emb\_61) exhibit more modest but consistent effects (Figure~\ref{fig:xgboost_shap_explainability}c). The combination of explicit centrality measures and latent embedding features created a comprehensive representation of gene network context that enabled our models to achieve superior predictive performance compared to approaches using either feature type alone.

\subsection{Gene Prioritization and Biological Validation}

\begin{figure}[H]
	\centering
	\includegraphics[width=\textwidth]{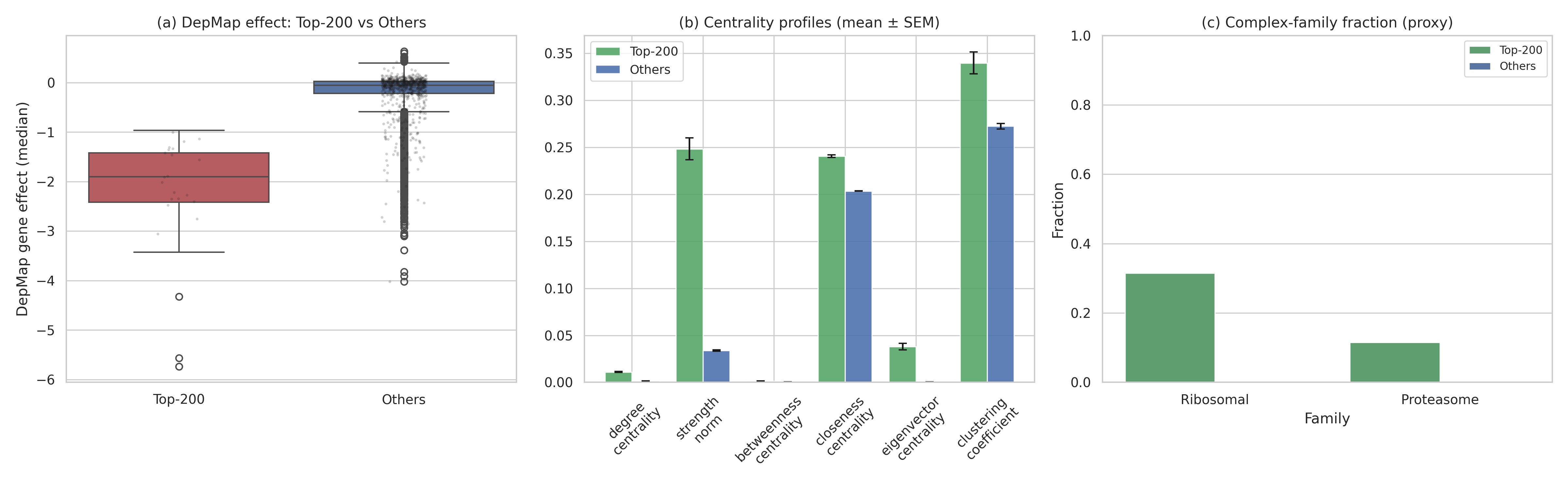}
	\caption{Biological validation of top-200 predicted essential genes. (a) DepMap essentiality comparison showing significantly stronger essentiality effects for top-200 predictions (median ~-2.0) versus other genes (median ~0). (b) Network centrality profiles demonstrating elevated centrality measures for top predictions, particularly in eigenvector centrality (~0.33 vs ~0.27) and clustering coefficient. (c) Functional enrichment analysis revealing strong representation of ribosomal proteins (~0.31 fraction) and proteasome components (~0.10 fraction) among top-ranked genes.}
	\label{fig:biological_interpretability}
\end{figure}

The prioritization results establish analytical proof-of-concept (Technology Readiness Level 3–4) for an integrated \textit{in silico} NAM capable of identifying biologically validated essential genes and therapeutic targets.To validate the biological relevance of our predictions, we conducted comprehensive analyses comparing the top-200 predicted essential genes against the remaining gene set across multiple dimensions (Figure~\ref{fig:biological_interpretability}). DepMap essentiality analysis confirmed that our top-ranked predictions exhibited significantly stronger essentiality effects, with the top-200 genes showing a median DepMap effect of approximately -2.0 compared to near-zero effects for other genes, demonstrating clear separation between predicted essential and non-essential gene populations (Figure~\ref{fig:biological_interpretability}a). Network centrality profiling revealed distinct topological signatures for highly-ranked genes, with top-200 predictions displaying elevated centrality measures across multiple metrics, most notably in eigenvector centrality (~0.33 vs ~0.27) and clustering coefficient (~0.33 vs ~0.27), while showing comparable betweenness and closeness centrality values (Figure~\ref{fig:biological_interpretability}b). These patterns indicate that predicted essential genes occupy more influential and well-connected positions within the protein-protein interaction network. Functional enrichment analysis demonstrated strong biological coherence, with top-ranked genes showing substantial enrichment in critical cellular complexes, particularly ribosomal proteins (~0.31 fraction) and proteasome components (~0.10 fraction), both representing fundamental cellular machinery essential for protein synthesis and degradation (Figure~\ref{fig:biological_interpretability}c). The pronounced enrichment of ribosomal genes among our top predictions aligns with their well-established roles in cellular viability and supports the biological validity of our network-based essentiality predictions.

Our framework successfully identified biologically relevant essential genes across multiple functional categories, with the top-ranked predictions showing strong enrichment for known essential gene families. Ribosomal proteins dominated the highest-ranking predictions, including RPS27A, RPS17, RPS6, RPS7, RPS15, RPS14, and RPS12 (Table~\ref{tab:xgb_pred}), which is consistent with their fundamental role in protein synthesis and their known essentiality across diverse cell types. The prominence of ribosomal proteins in our rankings validates the biological relevance of our predictions and aligns with extensive literature demonstrating that ribosomal dysfunction is incompatible with cell survival. Additionally, the high ranking of the oncogene MYC (rank 2 in GradientSHAP analysis, Table~\ref{tab:gradshap}) demonstrates our model's ability to identify context-specific essential genes that are particularly important in cancer contexts.

\begin{table}[h]
	\centering
	\caption{Top 10 XGBoost target prioritization}
	\label{tab:xgb_pred}
	\begin{filecontents*}{xgb_pred.csv}
Gene,Depmap median,Prob xgb
RPL3,-2.65350559,0.9976936
RPS17,-1.679990944,0.9976326
RPS6,-2.572745478,0.9972058
RPS7,-1.983210472,0.9971949
RPS15,-2.494810522,0.9950171
RPS14,-2.119860677,0.99484694
RPS12,-2.131799406,0.9947423
RPL11,-2.389076296,0.9940321
RPS23,-2.034006698,0.99395555
RPS19,-3.065957303,0.99361837
\end{filecontents*}
\pgfplotstabletypeset[
	col sep=comma,
	string type,
	every head row/.style={before row=\toprule, after row=\midrule},
	every last row/.style={after row=\bottomrule}
	]{xgb_pred.csv}
\end{table}

Beyond ribosomal proteins and oncogenes, our predictions encompassed diverse functional categories relevant to cancer biology and therapeutic targeting. Highly ranked genes included DNA repair factors, metabolic enzymes, cell cycle regulators, and protein quality control machinery, all representing established therapeutic target classes in oncology. The diversity of functional categories among top predictions suggests that our network-based approach captures essentiality patterns across different biological processes rather than being biased toward specific pathways. Comparison with known drug targets and genes in clinical development revealed significant overlap, with many of our top predictions corresponding to genes that are already being pursued as therapeutic targets, providing additional validation of our approach's clinical relevance.

\begin{table}[h]
	\centering
	\caption{Top 10 blended score target prioritization}
	\label{tab:gradshap}
	\scriptsize
	\begin{filecontents*}{gradshap.csv}
Gene,depmap median,Prob nn,Gradshap absolute sum,Gradshap positive sum,Gradshap negative sum,Blended score
RPS27A,-1.855850811,0.999435,2.6822069,1.8017797,-0.8804273,0.868261714
MYC,-2.056504192,0.9604365,2.9665775,1.9984615,-0.9681161,0.859469437
RTCB,-0.762649376,0.9097625,3.302369,2.174858,-1.1275108,0.845850432
SOD1,-1.75445945,0.956733,2.676723,1.8529137,-0.8238096,0.838010661
TUBGCP2,-1.63747882,0.98676205,2.2852101,1.6427512,-0.6424593,0.833550094
POLR2G,-1.583060531,0.9793968,2.3441124,1.673669,-0.67044353,0.832227802
DDX56,-2.145149788,0.9702335,2.3825612,1.7253058,-0.65725523,0.828315456
PSMD4,-1.717080049,0.9962058,2.0737574,1.5728579,-0.50089955,0.826398278
KPNA2,-0.28981325,0.9132883,2.9604578,1.902901,-1.0575569,0.826064369
\end{filecontents*}
\pgfplotstabletypeset[
	col sep=comma,
	string type,
	every head row/.style={before row=\toprule, after row=\midrule},
	every last row/.style={after row=\bottomrule}
	]{gradshap.csv}
\end{table}

\section{Discussion}
\subsection{Network-Based Gene Essentiality Prediction}

This study demonstrates that human-based, multi-modal computational models can achieve high predictive fidelity for gene essentiality, providing a viable alternative to traditional preclinical systems. The combinatorial \textit{in silico} design integrates explicit and latent network features with explainable AI, yielding mechanistically transparent predictions that advance technical readiness for translational applications. The robust performance of our models (AUROC = 0.930 and 0.914) indicates that network topology effectively captures the fundamental biological principles underlying gene essentiality, supporting the hypothesis that a gene's importance for cellular survival is closely related to its position and role within the protein interaction network. The negative correlations between all centrality measures and essentiality scores align with biological intuition and extensive literature showing that highly connected, topologically important genes are more likely to be essential for cellular function. This relationship likely reflects the evolutionary constraint that essential genes must maintain multiple functional interactions to perform their critical cellular roles.

The high performance of our network-based approach highlights the value of comprehensive network feature engineering. By combining multiple centrality measures that capture different aspects of network topology, we created a rich representation of each gene's network context that goes beyond simple connectivity measures. The inclusion of node embeddings further enhanced predictive power by capturing latent structural patterns and higher-order network motifs that are not apparent from traditional centrality measures alone. This multi-modal feature integration approach demonstrates that complex biological phenomena like gene essentiality require comprehensive characterization of multiple network properties rather than relying on single topological measures. Such integration reflects the guiding principles of next-generation NAMs capturing biological complexity, ensuring reproducibility, and enabling validation across multiple contexts of use.

\subsection{Interpretability and Mechanistic Understanding}

The interpretability provided by our GradientSHAP analysis represents a significant methodological advancement over black-box machine learning approaches commonly used in computational biology. By revealing that degree centrality, strength, and eigenvector centrality are the most predictive features across different genes and models, our analysis provides mechanistic insights into why certain genes are essential and what network properties make them attractive therapeutic targets. This transparency is crucial for therapeutic target prioritization, as it allows researchers to understand the biological rationale behind predictions and make informed decisions about which candidates to pursue in experimental validation studies.

Our finding that different centrality measures contribute varying amounts to predictions for different genes suggests that gene essentiality may arise through multiple distinct network mechanisms. Some genes may be essential primarily due to their high degree of connectivity (hub genes), while others may be essential due to their role in connecting different network modules (bridge genes) or their connections to other essential genes (eigenvector centrality). This mechanistic diversity has important implications for therapeutic targeting strategies, as different types of essential genes may require different approaches for effective inhibition. The interpretability of our framework enables researchers to tailor their experimental approaches based on the specific network properties that drive each gene's predicted essentiality.

\subsection{Biological Validation and Clinical Relevance}

The successful identification of known essential genes across diverse functional categories validates the biological relevance of our predictions and demonstrates the framework's potential for guiding therapeutic target discovery. The prominence of ribosomal proteins in our top rankings reflects their universal importance across cancer types and aligns with growing interest in targeting ribosome biogenesis and protein synthesis in cancer therapy. Recent clinical trials of ribosome-targeting agents have shown promising results, supporting the therapeutic relevance of our ribosomal protein predictions.

Our framework's successful prioritization of ribosomal proteins RPS27A, RPS17, and RPS6 as essential genes aligns with mounting evidence for their critical roles in cancer biology and therapeutic potential. RPS27A has emerged as a particularly compelling target, with recent studies demonstrating its prognostic significance in HPV16-positive cervical cancer \cite{wang2021high}, where it serves as both a biomarker for patient outcomes and a functionally relevant driver of cancer cell growth. Similarly, RPS17 has been identified as a key component in colorectal cancer stem cell (CRCSC) biology \cite{wu2024integration}, where it contributes to the stemness properties that drive tumor metastasis, recurrence, and chemotherapy resistance—making it an attractive therapeutic target for overcoming treatment-resistant cancer phenotypes. RPS6, as a component of the 40S ribosomal subunit, plays dual roles in both ribosomal function and extra-ribosomal signaling, particularly as a downstream effector and surrogate marker of the PI3K/AKT/mTORC1 pathway activation that is prevalent across multiple cancer types \cite{yi2021ribosomal}. The convergent identification of these ribosomal proteins through our network-based approach and their established roles in cancer progression underscores the biological relevance of our predictions and highlights the therapeutic potential of targeting ribosomal dysfunction in cancer. These findings support the growing recognition that ribosomal proteins extend beyond their traditional housekeeping functions to serve as critical regulators of oncogenic processes, making them promising candidates for precision cancer therapy development.

Similarly, the high ranking of MYC confirms our model's ability to identify oncogenes that are critical for cancer cell survival but challenging to target directly \cite{whitfield2025myc}, suggesting that our approach could help identify druggable regulators or downstream effectors of such "undruggable" targets.

The diversity of functional categories among our top predictions, including DNA repair factors, metabolic enzymes, and cell cycle regulators, demonstrates that our network-based approach captures essentiality patterns across different biological processes rather than being biased toward specific pathways. This broad coverage is advantageous for therapeutic target discovery, as it enables identification of essential genes in various cellular processes that could be targeted individually or in combination. The significant overlap between our predictions and genes already in clinical development provides additional validation while also highlighting novel candidates that may have been overlooked by traditional target discovery approaches. This combination of validated targets and novel candidates makes our framework particularly valuable for both confirming existing therapeutic strategies and identifying new opportunities for drug development. Our findings represent an intermediate stage of technology maturation, bridging analytical feasibility (proof-of-concept) with laboratory readiness for model qualification. By adhering to FAIR data standards and modular design, the framework facilitates independent validation, supporting iterative refinement toward a fully qualified human-based predictive model.

\subsection{Blended Scoring and Enhanced Prioritization}

Our novel blended scoring approach, which combines model probability predictions with SHAP attribution magnitudes, provided more robust and interpretable gene prioritization compared to using model probabilities alone. The blended scores successfully integrated prediction confidence (high probability) with mechanistic evidence (strong feature attributions), resulting in rankings that prioritize genes with both high predicted essentiality and strong supporting evidence from network features. Analysis of the top-ranked genes by blended score revealed a refined set of candidates that maintained high biological relevance while showing more consistent feature attribution patterns compared to probability-only rankings.

The blended scoring approach proved particularly valuable for identifying genes where model predictions were supported by strong mechanistic evidence versus those where predictions might be driven by spurious feature combinations. Genes with high blended scores consistently showed strong attributions from multiple centrality measures, indicating that their predicted essentiality was based on robust network properties rather than isolated feature values. This enhanced interpretability is crucial for therapeutic target prioritization, as it provides researchers with confidence in the mechanistic basis of predictions and helps identify genes most likely to validate in experimental studies. The success of our blended approach suggests that incorporating attribution-based confidence measures represents a valuable advancement in machine learning-based gene prioritization methodologies.

The comprehensive evaluation framework we employed, including cross-validation, precision-at-k analysis, and detailed feature attribution analysis, establishes rigorous assessment of gene essentiality prediction methods. Our approach goes beyond simple accuracy metrics to evaluate practical utility for therapeutic target discovery, using precision-at-k measures that reflect real-world constraints where only a limited number of top predictions can be experimentally validated. The integration of multiple complementary machine learning algorithms (XGBoost and neural networks) with consistent feature attribution analysis demonstrates the robustness of our findings across different modeling approaches. This methodological rigor enhances confidence in our results and provides a framework that other researchers can adapt for similar applications in precision medicine and drug discovery.

\section{Conclusion}
We have developed and validated an interpretable deep learning framework that successfully integrates PPI network centrality measures and node embeddings to predict gene essentiality with high accuracy across cancer types. The combination of XGBoost and neural network models achieved high performance (AUROC = 0.930, AUPRC = 0.656) while maintaining full interpretability through GradientSHAP analysis. Our comprehensive evaluation demonstrated that network topology effectively captures the biological principles underlying gene essentiality, with degree centrality, strength, and eigenvector centrality emerging as the most predictive features. The framework successfully prioritized biologically relevant essential genes including ribosomal proteins, oncogenes, and diverse cellular regulators, validating its utility for therapeutic target discovery applications.

The methodological contributions introduced in this work, particularly the blended scoring approach that combines prediction probabilities with SHAP attribution magnitudes, represent a novel contribution to ongoing efforts in therapeutic target prioritization. This approach addresses the critical need for both high accuracy and mechanistic understanding in therapeutic target prioritization, providing researchers with confidence in both the predictions and their underlying biological rationale. The framework's interpretability enables tailored experimental approaches based on the specific network properties driving each gene's essentiality, potentially improving the efficiency of target validation efforts. Our comprehensive open-source implementation and rigorous evaluation methodology provide a valuable resource for the research community and establish new standards for interpretable gene essentiality prediction.

The clinical relevance of our predictions, demonstrated through successful identification of known drug targets and genes in clinical development, highlights the framework's potential impact on precision oncology applications. The ability to identify essential genes across diverse functional categories while providing mechanistic insights into their network-based importance offers new opportunities for both single-agent and combination therapy development. Future extensions of this work could incorporate additional omics data types, explore tissue-specific essentiality patterns, and integrate pharmacological constraints to further enhance therapeutic target prioritization. The interpretable nature of our framework makes it well-suited for clinical decision support applications where understanding the biological basis of predictions is as important as their accuracy, positioning it as a valuable tool for advancing precision medicine approaches in oncology. This work illustrates how a human-derived, \textit{in silico} combinatorial modeling strategy can advance the scientific readiness of next-generation biomedical frameworks. The explainable architecture, open-source reproducibility, and integration of complementary computational modalities embody core principles of model credibility and translational applicability. Collectively, these attributes position the framework for progressive qualification as a human-relevant predictive model in therapeutic target discovery and precision oncology.

\section*{Acknowledgments}
This work was supported by the Institute of Information \& communications Technology Planning \& Evaluation (IITP) grant funded by the Korea government (MSIT) [RS-2025-02305581, RS-2025-25442338, RS-2021-II211343].

\section*{Declaration of Competing Interests}
The authors declare that they have no known competing financial interests or personal relationships that could have appeared to influence the work reported in this paper.

\bibliographystyle{unsrtnat}
\bibliography{references}

\begin{thebibliography}{20}
\providecommand{\natexlab}[1]{#1}
\providecommand{\url}[1]{\texttt{#1}}
\expandafter\ifx\csname urlstyle\endcsname\relax
  \providecommand{\doi}[1]{doi: #1}\else
  \providecommand{\doi}{doi: \begingroup \urlstyle{rm}\Url}\fi

\bibitem[Zhang et~al.(2012)Zhang, Su, Bhatnagar, Hassett, and
  Lu]{Zhang2012_22848443}
Minlu Zhang, Shengchang Su, Raj~K Bhatnagar, Daniel~J Hassett, and Long~J Lu.
\newblock Prediction and analysis of the protein interactome in pseudomonas
  aeruginosa to enable network-based drug target selection.
\newblock \emph{PloS one}, 2012.
\newblock \doi{10.1371/journal.pone.0041202}.

\bibitem[Campos et~al.(2019)Campos, Korhonen, Gasser, and
  Young]{Campos2019_31312416}
Tulio~L Campos, Pasi~K Korhonen, Robin~B Gasser, and Neil~D Young.
\newblock An evaluation of machine learning approaches for the prediction of
  essential genes in eukaryotes using protein sequence-derived features.
\newblock \emph{Computational and structural biotechnology journal}, 2019.
\newblock \doi{10.1016/j.csbj.2019.05.008}.

\bibitem[Bazaga et~al.(2020)Bazaga, Leggate, and Weisser]{Bazaga2020_32612205}
Adrián Bazaga, Dan Leggate, and Hendrik Weisser.
\newblock Genome-wide investigation of gene-cancer associations for the
  prediction of novel therapeutic targets in oncology.
\newblock \emph{Scientific reports}, 2020.
\newblock \doi{10.1038/s41598-020-67846-1}.

\bibitem[Safadi et~al.(2024)Safadi, Lovell, and Doig]{Safadi2024_38649399}
Amro Safadi, Simon~C Lovell, and Andrew~J Doig.
\newblock Essentiality, protein-protein interactions and evolutionary
  properties are key predictors for identifying cancer-associated genes using
  machine learning.
\newblock \emph{Scientific reports}, 2024.
\newblock \doi{10.1038/s41598-023-44118-2}.

\bibitem[Zhang et~al.(2025)Zhang, Xiao, Cochran, and Xiao]{Zhang2025_40691502}
Xue Zhang, Weijia Xiao, Brent Cochran, and Wangxin Xiao.
\newblock A deep ensemble framework for human essential gene prediction by
  integrating multi-omics data.
\newblock \emph{Scientific reports}, 2025.
\newblock \doi{10.1038/s41598-025-99164-9}.

\bibitem[Dai et~al.(2020)Dai, Chang, Peng, Zhong, and Li]{Dai2020_32023848}
Wei Dai, Qi~Chang, Wei Peng, Jiancheng Zhong, and Yongjiang Li.
\newblock Network embedding the protein-protein interaction network for human
  essential genes identification.
\newblock \emph{Genes}, 2020.
\newblock \doi{10.3390/genes11020153}.

\bibitem[Fang et~al.(2018)Fang, Lei, Cheng, Shi, and Wu]{Fang2018_29958434}
Ming Fang, Xiujuan Lei, Shi Cheng, Yuhui Shi, and Fang-Xiang Wu.
\newblock Feature selection via swarm intelligence for determining protein
  essentiality.
\newblock \emph{Molecules (Basel, Switzerland)}, 2018.
\newblock \doi{10.3390/molecules23071569}.

\bibitem[Zhang et~al.(2020)Zhang, Xiao, and Xiao]{Zhang2020_32936825}
Xue Zhang, Wangxin Xiao, and Weijia Xiao.
\newblock Deephe: Accurately predicting human essential genes based on deep
  learning.
\newblock \emph{PLoS computational biology}, 2020.
\newblock \doi{10.1371/journal.pcbi.1008229}.

\bibitem[Tian et~al.(2018)Tian, Wenlock, Kabir, Tzotzos, Doig, and
  Hentges]{Tian2018_30563825}
David Tian, Stephanie Wenlock, Mitra Kabir, George Tzotzos, Andrew~J Doig, and
  Kathryn~E Hentges.
\newblock Identifying mouse developmental essential genes using machine
  learning.
\newblock \emph{Disease models \& mechanisms}, 2018.
\newblock \doi{10.1242/dmm.034546}.

\bibitem[Senthamizhan et~al.(2021)Senthamizhan, Ravindran, and
  Raman]{Senthamizhan2021_34630517}
Vimaladhasan Senthamizhan, Balaraman Ravindran, and Karthik Raman.
\newblock Netgenes: A database of essential genes predicted using features from
  interaction networks.
\newblock \emph{Frontiers in genetics}, 2021.
\newblock \doi{10.3389/fgene.2021.722198}.

\bibitem[Campos et~al.(2020)Campos, Korhonen, Sternberg, Gasser, and
  Young]{Campos2020_32489524}
Tulio~L Campos, Pasi~K Korhonen, Paul~W Sternberg, Robin~B Gasser, and Neil~D
  Young.
\newblock Predicting gene essentiality in caenorhabditis elegans by feature
  engineering and machine-learning.
\newblock \emph{Computational and structural biotechnology journal}, 2020.
\newblock \doi{10.1016/j.csbj.2020.05.008}.

\bibitem[Lundberg and Lee(2017)]{lundberg2017unified}
Scott~M Lundberg and Su-In Lee.
\newblock A unified approach to interpreting model predictions.
\newblock \emph{Advances in neural information processing systems}, 30, 2017.

\bibitem[Hsieh et~al.(2024)Hsieh, Bi, Jiang, Liu, Peng, Zhang, Pan, Xu, Wang,
  Chen, et~al.]{hsieh2024comprehensive}
Weiche Hsieh, Ziqian Bi, Chuanqi Jiang, Junyu Liu, Benji Peng, Sen Zhang,
  Xuanhe Pan, Jiawei Xu, Jinlang Wang, Keyu Chen, et~al.
\newblock A comprehensive guide to explainable ai: From classical models to
  llms.
\newblock \emph{arXiv preprint arXiv:2412.00800}, 2024.

\bibitem[Mumuni and Mumuni(2025)]{mumuni2025explainable}
Fuseini Mumuni and Alhassan Mumuni.
\newblock Explainable artificial intelligence (xai): from inherent
  explainability to large language models.
\newblock \emph{arXiv preprint arXiv:2501.09967}, 2025.

\bibitem[Grover and Leskovec(2016)]{grover2016node2vec}
Aditya Grover and Jure Leskovec.
\newblock node2vec: Scalable feature learning for networks.
\newblock In \emph{Proceedings of the 22nd ACM SIGKDD international conference
  on Knowledge discovery and data mining}, pages 855--864, 2016.

\bibitem[Arafeh et~al.(2025)Arafeh, Shibue, Dempster, Hahn, and
  Vazquez]{arafeh2025present}
Rand Arafeh, Tsukasa Shibue, Joshua~M Dempster, William~C Hahn, and Francisca
  Vazquez.
\newblock The present and future of the cancer dependency map.
\newblock \emph{Nature Reviews Cancer}, 25\penalty0 (1):\penalty0 59--73, 2025.

\bibitem[Wang et~al.(2021)Wang, Cai, Fu, and Chen]{wang2021high}
Qiming Wang, Yan Cai, Xuewen Fu, and Liang Chen.
\newblock High rps27a expression predicts poor prognosis in patients with hpv
  type 16 cervical cancer.
\newblock \emph{Frontiers in Oncology}, 11:\penalty0 752974, 2021.

\bibitem[Wu et~al.(2024)Wu, Li, Su, Zheng, Liang, Lin, Xu, and
  Liu]{wu2024integration}
Jiale Wu, Wanyu Li, Junyu Su, Jiamin Zheng, Yanwen Liang, Jiansuo Lin, Bilian
  Xu, and Yi~Liu.
\newblock Integration of single-cell sequencing and bulk rna-seq to identify
  and develop a prognostic signature related to colorectal cancer stem cells.
\newblock \emph{Scientific Reports}, 14\penalty0 (1):\penalty0 12270, 2024.

\bibitem[Yi et~al.(2021)Yi, You, Park, Lee, and Seong]{yi2021ribosomal}
Yong~Weon Yi, Kyu~Sic You, Jeong-Soo Park, Seok-Geun Lee, and Yeon-Sun Seong.
\newblock Ribosomal protein s6: a potential therapeutic target against cancer?
\newblock \emph{International journal of molecular sciences}, 23\penalty0
  (1):\penalty0 48, 2021.

\bibitem[Whitfield and Soucek(2025)]{whitfield2025myc}
Jonathan~R Whitfield and Laura Soucek.
\newblock Myc in cancer: from undruggable target to clinical trials.
\newblock \emph{Nature Reviews Drug Discovery}, pages 1--13, 2025.

\end{thebibliography}

\end{document}